\documentclass[prb,twocolumn,showpacs]{revtex4}
\usepackage{graphicx}
\usepackage{epsfig}
\usepackage{mathrsfs}
\usepackage{amsmath,amssymb}

\begin{document}
 \title{From Coulomb blockade to the Kondo regime in a Rashba dot}
\author{Rosa L\'opez}
\affiliation{Departament de F\'{i}sica, Universitat de les Illes Balears,
 E-07122 Palma de Mallorca, Spain}
\author{David S\'anchez}
\affiliation{Departament de F\'{i}sica, Universitat de les Illes Balears,
  E-07122 Palma de Mallorca, Spain}
\author{Lloren\c{c} Serra}
\affiliation{Departament de F\'{i}sica, Universitat de les Illes Balears,
  E-07122 Palma de Mallorca, Spain}
\affiliation{Institut Mediterrani d'Estudis Avan\c{c}ats IMEDEA (CSIC-UIB), 
E-07122 Palma de Mallorca, Spain}
\date{\today}
\begin{abstract}
We investigate the electronic transport in a quantum wire with localized 
Rashba interaction.
The Rashba field forms quasi-bound states which couple to the continuum states
with an opposite spin direction. The presence of this Rashba dot causes 
Fano-like antiresonances and dips in the wire's linear conductance. The
Fano lineshape arises from the interference between the direct transmission channel
along the wire and the hopping through the Rashba dot. Due to the confinement, we
predict the observation of large charging energies in the local Rashba region
which lead to Coulomb-blockade effects in the transport properties of the
wire. Importantly, the Kondo regime can be achieved with a proper tuning
of the Rashba interaction, giving rise to an oscillating linear conductance for a fixed
occupation of the Rashba dot.  
\end{abstract}
\maketitle
\section{Introduction}
The high degree of functionality of spin-based
electronic devices has attracted much attention
due to promising applications in quantum
and classical computation.~\cite{Wolf01} Use of spins to store and carry
classical information has been proven to be both faster and less power
consuming than the conventional technology based on the control of the charge.
Besides, spin-1/2 systems are genuine two-level systems
and, therefore, natural building blocks for quantum computation.~\cite{Loss98}
These features have motivated the emergence of an exciting area of research
termed \emph{spintronics}.

Two-dimensional (2D) semiconductors are appropriate materials to accomplish
spintronic applications since they offer the possibility of electric control
of spins via tunable \emph{spin-orbit} interactions. The tuning is achieved
with electric fields induced by external gates coupled to the semiconductor.~\cite{nit97}
A prominent contribution to spin-orbit effects in 2D electron gases of
narrow-gap semiconductors (typically, InAs materials) is the Rashba
interaction.~\cite{ras60} This is caused by an asymmetry in the potential
defining the quantum well in the direction perpendicular to the 2D electron
gas. The control of the Rashba coupling strength
opens the possibility of investigating 2D systems with spatially
modulated Rashba fields such as those having constant Rashba strength 
in a semiplane \cite{kod}, a stripe \cite{val1} or an island.
\cite{xu05,Pal06,egu02,val2,llorens} Less known are the effects of a local
Rashba field on the electron-electron interaction. Our aim is to examine how
sensitive the charging energy of a quasi-1D Rashba region is to changes of the
spin-orbit strength and the consequences in the transport properties.

The Rashba Hamiltonian reads
\begin{equation}
\label{rashbaH}
\mathcal{H}_{\rm R}=\frac{1}{2\hbar}( \{\alpha,
p_{y}\}\sigma_{x}-\{\alpha,p_{x}\}\sigma_{y})\,,
\end{equation}
where $\alpha$ is the strength while $\vec{p}$ and $\sigma_{x,y,z}$ are the 2D
momentum operator and the Pauli matrices, respectively. The anticonmutators in
Eq.~(\ref{rashbaH}) ensure a Hermitian $\mathcal{H}_{\rm R}$ when
$\alpha$ is spatially nonuniform. For strictly 1D systems the precession term
in Eq. (\ref{rashbaH}), $\{\alpha, p_{x}\}\sigma_{y}$ for transport along $x$,
leads to the formation of bound states.~\cite{llorens} These bound states
acquire a finite lifetime in quasi-1D systems where the term
$\{\alpha,p_{y}\}\sigma_{x}$ couples adjacent subbands with opposite spin
directions.~\cite{mir01} In a quantum wire these subbands arise from the
parabolic confinement of the 2D gas. Therefore, a local Rashba interaction in
a quantum wire has two main effects, namely \emph{(i) it forms bound states}
in each subband and \emph{(ii) it broadens these bound states due to coupling
  with adjacent subbands}. Hereafter, we refer to the quasibound states as \emph{Rashba dots}.~\cite{llorens}

In this work we are interested in the transport properties of a quantum
wire with a local Rashba interaction in the
presence of Coulomb repulsion. As we have partially anticipated,
the physical scenario is governed by the
interference between a direct (nonresonant) transmission channel with
a path that passes through the Rashba dot,
leading to the formation of \emph{Fano} resonances~\cite{fano,llorens}
as shown in Fig. \ref{fig:1}(a).
Due to the confinement of the electron motion within the Rashba dot
our calculations show that the Coulomb 
interaction becomes in fact very large, see Fig. \ref{fig:1}(b).
Thus, it is possible to observe charging
effects where the transport through the Rashba dot is governed by Coulomb
blockade. Furthermore, a proper combination of
gates leads to a very strong coupling of a singly occupied Rashba dot to the
continuum states. In this case and when the Coulomb interaction is
sufficiently large the \emph{Kondo effect}~\cite{hew93,theory} takes place at very low
temperatures. Then, the localized spin in the Rashba dot
forms a many-body singlet state with the continuum electron spins and,
hence, is screened.  This occurs at energies
$k_BT$ lower than the Kondo energy scale $k_BT_K$, which is the binding energy
of the many-body singlet state.~\cite{hew93,theory}  It manifests itself as a
quasi-particle resonance, namely, a peak of width $k_BT_K$ of the local
density of states (LDOS) at the Fermi energy ($E_F$).  Experiments
have been able to observe Kondo effects in quantum
dots,~\cite{experiment} its most remarkable signature being the
\emph{unitary limit} of the linear conductance at $k_BT=0$, i.e., $\mathcal{G}_{0}=2e^2/\hbar$.~\cite{theory} In our
case, the Kondo resonance that forms in the Rashba dot destructively
interferes with the nonresonant transmitting mode giving rise to an
oscillatory $\mathcal{G}_{0}$.

This work considers the transport properties of a quantum
wire with a localized Rashba region including charging and correlation
effects. Section II presents the model of Hamiltonian developed to describe
this system. Section III is devoted to investigate the linear conductance
across the quantum wire with the localized Rashba interaction. The results
derived from our calculations are shown in Section III. Finally, the main conclusions of our work are summarized in Section IV. 

\section{Model} 

In this section we discuss the model Hamiltonian that describes a
quantum wire with a localized Rashba interaction. 
As emphasized, the Rashba interaction plays
the role of both (i) the attractive potential
and (ii) the coupling to the continuum states. Each subband couples with at
least one bound state (Rashba dot) splitting off the higher subband. For simplicity we consider an energy range in the first plateau,
$\varepsilon_1<E<\varepsilon_2$ with $\varepsilon_\nu=(\nu-1/2)\hbar\omega_0$ 
being the energy of the transversal modes ($\omega_0$ defines
the confinement strength of the quasi-1D system).
The bound state energy $\epsilon_0=\varepsilon_{2}+\epsilon_{b}$ lies within
the first plateau because of its negative binding energy $\epsilon_b$. 
Thus, the first subband consists of a continuum of states and
the second subband provides the bound state which is coupled 
to the first subband states with opposite spin. As a consequence, the
transport can occur through two different paths, (i) via the quasi-bound state
(through the Rashba dot described by an
Anderson-like Hamiltonian) or (ii) via a
nonresonant path through the continuum states of the first subband.
The Hamiltonian reads:
\begin{multline}\label{model}
\mathcal{H} = 
\sum_{\alpha k\sigma} 
\epsilon_{\alpha k}\,
c_{\alpha k\sigma}^\dagger   
c_{\alpha k\sigma}^{\rule[-1pt]{0pt}{6pt}} +  
\sum_{k\sigma} (W e^{i s_\sigma\varphi} 
c^\dagger_{L k\sigma} 
c_{R k \sigma}^{\rule[-2pt]{0pt}{6pt}} + {\rm H.c.}) \\
+ \sum_{\sigma} \epsilon_0\, 
d^\dagger_\sigma 
d^{\rule[-1pt]{0pt}{6pt}}_\sigma + 
U n_{\sigma} n_{\bar\sigma}+
\sum_{\alpha k\sigma} (V_{0}  
c^\dagger_{\alpha k\sigma} 
d_{\bar{\sigma}}^{\rule[-1pt]{0pt}{6pt}}
+ {\rm H.c.} )\; .
\end{multline}
We choose the spin quantization axis along the Rashba field (the $y$-axis for
propagation along $x$). $d^\dagger_{\sigma}$ creates an electron with spin
$\sigma=\uparrow,\downarrow$ in the Rashba dot ($n_{\sigma}=d^\dagger_\sigma 
d^{\rule[-1pt]{0pt}{6pt}}_\sigma$ is the Rashba dot occupation per spin $\sigma$) and $c^\dagger_{L(R) k\sigma}$ creates a continuum electron with $k$
wavevector and spin $\sigma$ in the left (right) side of the Rashba
dot. Both electronic states are coupled via the hopping amplitude $V_0$.
We note that this hopping originates from the Rashba interaction only,\cite{llorens}
and that this qualitatively differs from the coupling via tunnel barriers
as in conventional quantum dot models.
We define the signs $s_{\uparrow,\downarrow}=\pm 1$ and $\bar\sigma$ indicates
reverse spins. $U$ is the on-site Coulomb interaction in the Rashba
dot. Depending on the strength of $U$, which we calculate below, charging and
correlation effects can be present. 
We note that the Hamiltonian (\ref{model}) presents a
great similarity with that describing the electronic transport in a closed
Aharanov-Bohm (AB) interferometer with a quantum dot in one of its
arms.~\cite{Yac95,Butt95,bulka01,konig01}
Nevertheless, there are two important differences. First, the phases of the hopping amplitudes $W$
in Eq.\ (\ref{model}) depend on the spin index.
Such spin-dependent phases have been shown to give rise to spin
polarization in interferometer setups.~\cite{sun05}
Second, each hopping process through the dot is associated
with a spin-flip event. Spin-flip interactions have been recently considered 
in strongly correlated quantum dots~\cite{sf} but mainly dealing with
intradot spin-flip scattering. Spin-flip assisted tunneling has
received much less attention.~\cite{lin05,sun05}

In the usual
Anderson Hamiltonian, spin is conserved during tunneling and this leads
to Kondo correlations at low temperatures. 
We now show that, despite the fact that Eq.~(\ref{model}) involves spin-flip
hopping due to the Rashba interaction, the Kondo effect
also manifests itself in this system. For simplicity, let us neglect the
nonresonant path in the following discussion ($W=0$). We first perform a
Schrieffer-Wolff transformation to the Rashba dot Hamiltonian. The resulting
Hamiltonian,
\begin{equation}\label{kondo2}
\mathcal{H'}_{K}=J_{\alpha\beta}(S^{x} s_{\alpha\beta}^{x}-S^{y} s_{\alpha\beta}^{y}-S^{z} s_{\alpha\beta}^{z})\,,
\end{equation}
is equivalent to the
usual Kondo Hamiltonian 
\begin{equation}\label{kondo3}
\mathcal{H}_{K}=\sum_{\alpha\beta}J_{\alpha\beta}
\vec{S}\cdot \vec{s}_{\alpha\beta}\,,
\end{equation} 
where the spin of the Rashba dot $\vec{S}$ interacts
antiferromagnetically ($J_{\alpha\beta}>0$) with the spin density of the
delocalized electrons
$\vec{s}_{\alpha\beta}=\psi^{\dagger}_{\alpha\nu}\vec{\sigma}_{\nu\mu}\psi_{\beta\mu}$
($\psi_{\alpha \nu}=\sum_{k} c_{\alpha k\nu}$ are the conduction field
operators at the Rashba dot site).  To see this, we can perform a unitary
transformation to convert  $\mathcal{H'}_{K}$ into $\mathcal{H}_{K}$, namely,
we perform a rotation of $\pi$ around the $x$
axis in the localized spin space keeping the delocalized spins unchanged,
i.e., 
\begin{equation}
S^{y(z)}\dashrightarrow -S^{y(z)},\,\,\,\,S^{x}\dashrightarrow S^{x}\,.
\end{equation}
In this way we transform $\mathcal{H'}_{K}$ into $\mathcal{H}_{K}$.
\\
Physically, the system we consider is characterized
by just two parameters: $\alpha$, the intensity of the Rashba field, 
and $\ell$, the length where this is active in the quantum wire [see inset 
of Fig.\ \ref{fig:1}(b)]. Importantly, the parameters
governing the Hamiltonian~(\ref{model}), $U\equiv U(\alpha, \ell)$,
$V_{0}\equiv V_{0}(\alpha, \ell)$, $W\equiv W(\alpha, \ell)$, and
$\varphi(\alpha,\ell)$ can be externally
controlled by changing both $\alpha$ and $\ell$.
The Rashba phase $\varphi= k_\alpha\ell$ with $k_\alpha= m\alpha/\hbar^2$ 
is the total phase gained by an
electron traveling from the left to the right side of the Rashba dot.

\section{Linear Conductance} 

Once we have established, in the previous
section, that in a quantum wire with a localized Rashba
interaction Kondo physics can arise, now we derive the transport properties
for such a system. 

The electrical current is derived from
the time derivative of the quantum occupation of the conduction electrons on
the left side of the Rashba dot: 
\begin{equation}
I_{L}=e d \langle N_{L}\rangle
/dt=(ie/\hbar)\langle[\mathcal{H}, N_L]\rangle\,,
\end{equation}
where $N_{L}=\sum_{k\sigma}
c^\dagger_{L k\sigma}c_{L k\sigma}$.  In terms of the nonequilibrium Green
functions $G_{LR \sigma}^{<}=i\langle c_{R k\sigma}^\dagger
c_{L k\sigma}\rangle$ and $G_{d\bar{\sigma}, 
L\sigma}^{<}=i\langle c_{L k\sigma}^\dagger d_{\bar\sigma} \rangle$ the current reads
\begin{equation}
I_{L}\!=\frac{2e}{h}\mathcal{R}e\sum_{k\sigma} \!\!\int\! d\epsilon \!\!\left[W e^{i s_\sigma\varphi}
 G_{LR\sigma}^{<}(\epsilon) \!+ \!V_{0} G_{d\bar{\sigma}, L\sigma}^{<}(\epsilon)\right]\!\!.
\end{equation}
After some algebra, the current traversing the system can be written in
terms of the transmission as 
\begin{equation}
I=(e/h)\sum_\sigma\int d\epsilon \mathcal{T}_{\sigma}(\epsilon)
[f_{L}(\epsilon)-f_{R}(\epsilon)]\,,
\end{equation}
where  $f_{L(R)}$ is the left (right)
Fermi function and the transmission reads
\begin{multline}\label{transmission}
 \mathcal{T}_{\sigma}(\epsilon)=T_{b}+2\sqrt{T_{b} R_{b}}\cos(s_\sigma\varphi) \tilde{\Gamma}
 \mathcal{R}e G^{r}_{d\bar{\sigma}}(\epsilon) 
\\-\tilde{\Gamma}\left\{
 \left[1-T_{b}\cos^2(s_\sigma\varphi)\right]-T_{b} \right\} \mathcal{I} m G^{r}_{d\bar{\sigma}}(\epsilon)\,.
\end{multline}
In Eq.\ (\ref{transmission}),
\begin{equation}
T_{b}=4 x^2 /(1+ x^2)^2,\,\, (R_b=1-T_b)\,,
\end{equation}
 is the background transmission with
$x=\pi W \rho $ ($\rho$ is the DOS of the conduction electrons), and
$\tilde{\Gamma}=\Gamma/(1+x^2)$ with $\Gamma=\pi V_{0}^2\rho$. $G^{r}_{d\sigma}$ is the retarded Green function for the Rashba dot. This Green function is
determined by calculating its self-energy which takes into account the
single-particle self-energy due to tunneling and the many-body effects
enclosed in the interacting self-energy $\Sigma_{\rm int}(\epsilon)$:
\begin{equation}
G^{r}_{d\sigma}(\epsilon)=\frac{1}{\epsilon-\epsilon_{0\sigma}+i\tilde\Gamma-\Sigma_{\rm
    int}(\epsilon)}\,.
\end{equation}
Notice that Eq.~(\ref{transmission}) is equivalent to the transmission
obtained in Refs.~\onlinecite{bulka01,konig01} for the closed AB interferometer since
\begin{equation}
\cos(s_\sigma\varphi)=\cos(\varphi)\,,\,\,\,\,G_{d\sigma}^r=G_{d\bar\sigma}^r\,,\end{equation}
whenever the spin degeneracy is not broken. In linear response the conductance
is just 
\begin{equation}
\mathcal{G}_{0}\equiv \lim_{V\rightarrow 0} \frac{dI}{dV}=
\frac{2e^2}{h}\mathcal{T}(E_F)\,.
\end{equation}
The spin-degenerate transmission 
at the Fermi energy $\mathcal{T}(E_F)$ can be
written in an extended Fano form as
\begin{equation}\label{Tfano}
\mathcal{T}(E_F)=T_{b}\frac{|\xi+q|^2}{\xi^2+1}\,,
\end{equation}
with a \emph{complex Fano parameter}
\begin{equation}
q=\sqrt{\frac{1}{T_{b}}}\left(\sqrt{R_b}\cos\varphi+i\sin\varphi\right)\,,
 \end{equation}
and
\begin{equation}
\xi=\frac{2\left[E_F-\epsilon_0-\mathcal{R}e \Sigma_{\rm int}(E_F)\right]}{\tilde\Gamma-\mathcal{I}m \Sigma_{\rm int}(E_F)}\,.
 \end{equation}
Importantly, $q$ is
complex even though the Rashba interaction is time-reversal invariant. This is
in clear contrast with the Aharanov-Bohm case where the complex Fano factor is
usually attributed to the broken time-reversal symmetry by the magnetic
field.~\cite{kob} Therefore, further investigation should be necessary to clarify the
general conditions for the appearance of a complex Fano factor.~\cite{us}

\subsection{Zero temperature: Friedel Langreth sum rule}
At $k_B T=0$ the quasiparticles in an interacting system (a
Fermi liquid) possess an infinite life-time, meaning that the imaginary part of the interacting
self-energy vanishes, i.e.,  $\mathcal{I}m\Sigma_{\rm int}(E_F)=0$.
According to this,  the Friedel-Langreth sum rule~\cite{lan66} relates
the scattering phase shift $\delta$, picked up by the electrons when hopping,
with the total quantum occupation $\langle n \rangle=\sum_{\sigma}\langle n_{\sigma} \rangle
$ of the interacting system
\begin{equation}
\delta=\frac{\pi \langle n \rangle}{2}\,.
\end{equation} 
The quantum occupation for the interacting system at $T=0$ can be easily obtained in terms of $\xi$
\begin{equation}
 \langle n \rangle=\frac{2}{\pi}{\rm cot^{-1}}\xi\,.
\end{equation} 
As a consequence, the transmission at the Fermi level,
$\mathcal{T}(E_F)$, can be cast in terms of $\delta$ only. Moreover, in
the pure Kondo regime the dot level is renormalized
by the interactions and pinned at the Fermi energy, i.e.,
\begin{equation}
\epsilon_0+\mathcal{R}e \Sigma_{\rm int}(E_F)=E_F\,,
\end{equation}
giving rise to the Kondo
resonance. As a result, the occupation per spin in the pure Kondo regime does not fluctuate
and remains constant: $\langle n_\sigma \rangle =1/2$. This fact implies that the
linear conductance has a simple form:~\cite{konig01}  
\begin{equation}
\mathcal{G}_{0}=\frac{2e^2}{h}\left(1-T_{b}\cos^2\varphi\right)\,.
 \end{equation}
Unlike the AB interferometer where $\varphi$ is constant, now the phase
changes with the Rashba intensity. This means that while the Kondo regime for
the AB interferometer shows a \emph{globally} reduced linear conductance,
independently of the gate position,~\cite{konig01,bulka01,Sato} for the Rashba dot the hallmark of the
Kondo effect consists of an {\em oscillating} $\mathcal{G}_0$ with $\alpha$,
due to the variation of $\varphi$. 
\begin{figure}
\centering
\includegraphics*[width=90mm]{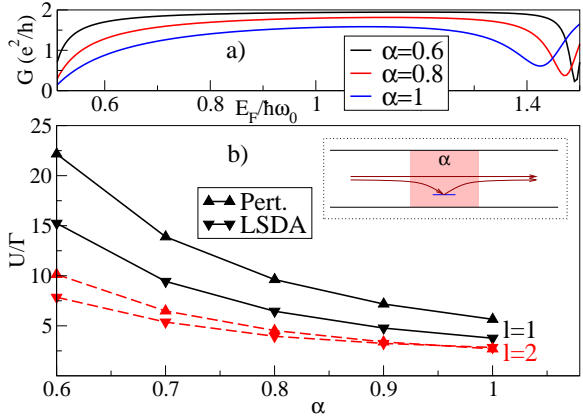}
\caption{(a) (Color online) (a) Linear conductance $\mathcal{G}_{0}$ of a noninteracting quantum wire
  as a function of the Fermi energy $E_F$ (in units of $\hbar\omega_0$)
  with $\ell=1$ (in units of $\ell_0=\sqrt{\hbar/m\omega_0}$) and different Rashba strengths
  $\alpha$ (in units of $\hbar\omega_0\ell_0$). A dip forms close to the
onset of the second plateau. It follows from the formation of a quasibound
state which couples to the conduction subband [see Inset in (b) for a sketch],
giving rise to a Fano lineshape in $\mathcal{G}_{0}$. Both the position and
broadening of the dip are tuned with $\alpha$. (b) Charging energy $U$
normalized to the coupling broadening $\Gamma$ as a function of $\alpha$ for different $\ell$. Data sets are obtained using a perturbative expression and LSDA for $e^2/\varepsilon l_0=\hbar\omega_0$.
}
\label{fig:1}
\end{figure}
\subsection{Finite temperature}
For finite $k_BT$ the Friedel-Langreth sum rule does not hold and the Green function for
the Rashba dot has to be derived explicitly. For that purpose we make use of the
equation-of-motion technique.~\cite{Lacroix} This method is based on
differentiating the dot Green function with respect to time generating higher-order Green function's that are approximated following a truncation scheme. Here we choose Lacroix's approximation~\cite{Lacroix} which goes beyond mean-field theory to include Kondo correlations,
evaluating the self-energies in the wide-band limit:
\begin{multline}\label{GF}
G_{d\sigma}^{r}(\epsilon)=\frac{1-\langle n_{\bar\sigma}\rangle}{\epsilon-\epsilon_{0}-\Sigma_{0}-\Sigma^{(1)}_{\rm
    int}}\\
+\frac{\langle n_{\sigma}\rangle }{\epsilon-\epsilon_{0}-U-\Sigma_{0}-\Sigma^{(2)}_{\rm int}}\,,
\end{multline}
where 
\begin{eqnarray}
&&\Sigma_{0}=-i\tilde{\Gamma}\left(1-ix\cos\varphi\right)\,,\\
&&\Sigma^{(1)}_{\rm int}(\epsilon)=\frac{U\bar\Upsilon(\epsilon)}{\epsilon-\epsilon_0-U-\Sigma_{0}-\bar\Sigma_1}\,,\\
&&\Sigma^{(2)}_{\rm int}(\epsilon)=-\frac{U\left[\Sigma_{1}-\bar\Upsilon(\epsilon)\right]}{\epsilon-\epsilon_0-\Sigma_{0}-\bar\Sigma_1}\,.
\end{eqnarray}
By defining 
\begin{equation}
\Upsilon(\epsilon)=\Psi\left(\frac{1}{2}-i\beta\frac{\epsilon-(2\epsilon_0+U)}{2\pi }\right)-\Psi\left(\frac{1}{2}-i\beta\frac{\epsilon}{2\pi }\right)-i\pi\,,
\end{equation} 
where $\Psi$ is the digamma function, $D$ is bandwidth for the DOS in the
leads, and $\beta=1/k_BT$, we have
$\bar\Upsilon(\epsilon)=(\tilde\Gamma/\pi)\Upsilon(\epsilon)\left(1+\pi
x\cos\varphi\right)$. Here $\bar\Sigma_1=-2i\tilde\Gamma\left(1+
x\cos\varphi\right)$. Using this result for the Rashba dot Green function in Eq.~(\ref{transmission}) we can calculate the linear conductance and explore both the Coulomb blockade and the Kondo regimes.\begin{figure}
\centering
\includegraphics*[width=90mm]{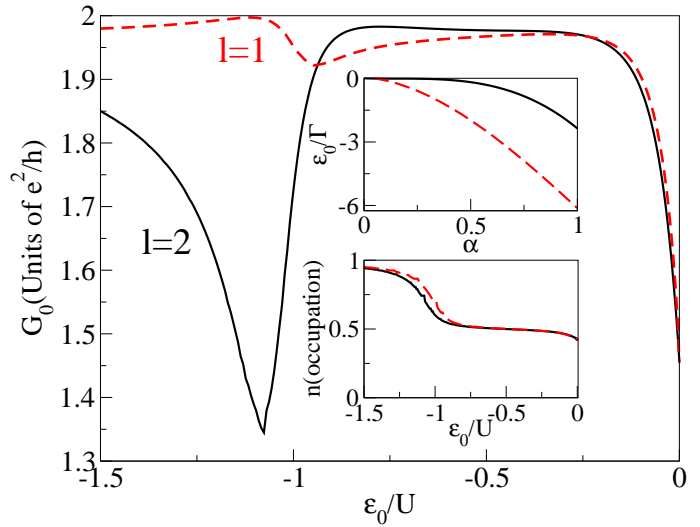}
\caption{(Color online) $\mathcal{G}_0$ in the Coulomb-blockade regime
for $x=0.9$, $\ell=1,2$ and $k_BT=0.1\Gamma$ (we set $E_F=0$). For better comparison, the
charging energy remains unchanged with $\alpha$ and $\ell$: $U=7\Gamma$.
Upper inset: Nonlinear dependence of $\epsilon_0$ with $\alpha$.
Lower inset: Dot occupation versus gate voltage.}
\label{fig:2}
\end{figure}

\section{Results} For a noninteracting quantum wire ($U=0$) with a local
Rashba region (see inset in Fig.~\ref{fig:1}) $\mathcal{G}_{0}$ versus $E_F$
is shown in Fig.~\ref{fig:1}(a) for three different values of $\alpha$. Here, 
the formation of the Rashba dot can be seen in the
appearance of an antiresonance located at the end of the first
plateau.~\cite{llorens} Our next question is about the strength of $U$ and how
this can affect the previous results. With a finite $U$ the transport through
a singly occupied quantum dot is
blockaded due to the energy cost required to charge the dot with another
electron. To overcome this situation and restore the transport the band bottom
of the dot must be pulled down by
applying an external gate.
For the present case of a Rashba dot the same effect is accomplished by increasing
$\alpha$. This is the so-called Coulomb blockade regime that arises whenever
$U\gg\tilde\Gamma\gg k_B T$. At very low temperatures $k_BT\ll k_BT_K$ the Kondo effect occurs because of a strong $U$
in a quantum dot strongly coupled to the leads.~\cite{hew93} In this
regime the formation of a singlet state between the itinerant electrons and
the localized electron in the quantum dot opens a channel of unitary transmission.~\cite{theory}
In our system we check that these two regimes can be reached for typical
values of $\alpha$. Figure~\ref{fig:1}(b) displays a calculation of
$U/\Gamma$ obtained from two different approaches. First, we use a perturbative expression, 
\begin{equation}
U_{\rm pert}=-\frac{e^2}{\varepsilon}\int d\vec{r}d\vec{r}' \frac{|\phi_{0}(\vec{r})|^2
 |\phi_{0}(\vec{r}')|^2}{|\vec{r}-\vec{r}'|}\,,
\end{equation}
 where
$\phi_{0}(\vec{r})$ is the wavefunction of the induced bound state by the Rashba
interaction and $\varepsilon$ is the material permittivity. In this way $U$ is the bare Coulomb interaction which does not
include screening effects due to the presence of the rest of charges. To
perform a more realistic calculation of $U$ taking into account screening
effects we employ the standard local spin-density
approximation (LSDA).~\cite{Rei02}. In spite of the different approaches
used to determine $U$ one clearly sees that both calculations lead to similar
results for $U$. Indeed $U$ is quite large. This means that the effects of a finite $U$ cannot be ignored
and, as a result, this changes dramatically the transport properties of this
system. Notice that for $\alpha\approx 0$, our calculation for $U$ fails since the wave function in the Rashba dot is
now well extended along the quantum wire and not only in the Rashba
region. In that case the onsite interaction becomes very
weak $U/\tilde\Gamma\rightarrow 0$ and the noninteracting
theory~\cite{llorens} can be safely applied. 
Hereafter, we consider only situations where the Rashba coupling takes values
ranging between $\alpha\approx 0.5-1$ [see Fig.~\ref{fig:1}(b)]. 

First, we analyze the effect of a finite $U$ that carries
this system into the Coulomb blockade regime for $k_B T>k_BT_K$, and the Kondo
regime when $-U+E_F+\tilde\Gamma<\epsilon_0<E_F-\tilde\Gamma$ and $k_B T\ll
k_BT_K$. Typically, in a conventional quantum dot the  charging energies for
observing correlation effects are in the range of 0.5-1meV with
tunneling couplings lying between 100$\mu$~eV and 300$\mu$~eV.~\cite{experiment}
This implies the ratio $U/\Gamma\approx 3-10$ which can be easily achieved in our
device [see Fig.~\ref{fig:1}(b)].

Figure~\ref{fig:2} displays $\mathcal{G}_{0}$ versus 
$\epsilon_{0}/U$ for two different values of $\ell$
calculated in the Hartree-Fock approximation,~\cite{Lacroix} where
$\Sigma^{(1)}_{\rm int}(\epsilon)=\Sigma^{(2)}_{\rm int}(\epsilon)=0$ in Eq.~(\ref{GF}), neglecting in this way
correlation effects in the leads. As we mentioned before, $\alpha$ plays
the role of the gate voltage and controls the level position
$\epsilon_{0}/\Gamma$ that depends at the same time on $\ell$.
The upper inset of Fig.~\ref{fig:2} shows such
dependence for $\ell=1$ and $\ell=2$. At these level positions (or
their corresponding values of $\alpha$ and $\ell$) $\mathcal{G}_{0}$
consists of two asymmetric Fano resonances located approximately at 
\begin{eqnarray}
&&\epsilon^{(1)}_{0}\approx -\Sigma_0=(\tilde\Gamma/2) x\cos\varphi\nonumber\,,\\
&&\epsilon^{(2)}_{0}\approx
-U-\Sigma_0=-U+(\tilde\Gamma/2)x\cos\varphi\,.
\end{eqnarray}
These two antiresonances (the resonance at $\epsilon_0\sim E_F$ is seen in
Fig.~\ref{fig:2} only partially
because $\epsilon^{(1)}_{0}>0$ for that value of $\epsilon_0$)
arise from the interference between the nonresonant path with the two resonances corresponding to the degenerate points in which the transport
through the Rashba dot is energetically allowed. At these points the dot
occupation $\langle n_{\sigma}\rangle $ fluctuates from $0\rightarrow 1/2$ for values of the
gate close to $\epsilon^{(1)}_{0}$ and from $1/2\rightarrow 1$ at values close
to $\epsilon^{(2)}_{0}$, as seen in the lower inset to Fig.~\ref{fig:2}.
Remarkably, the position of the Fano resonances is not very sensitive to a
change in $\ell$ since this only changes the Rashba phase $\varphi=k_\alpha\ell$.
We recall that $U\gg \tilde\Gamma$ and that, typically, for
quantum wires $x\approx 1$ ($T_{b}\approx 1$). 
However, we anticipate that in the Kondo regime a change in $\ell$ modifies dramatically $\mathcal{G}_{0}$.

\begin{figure}
\centering
\includegraphics*[width=90mm]{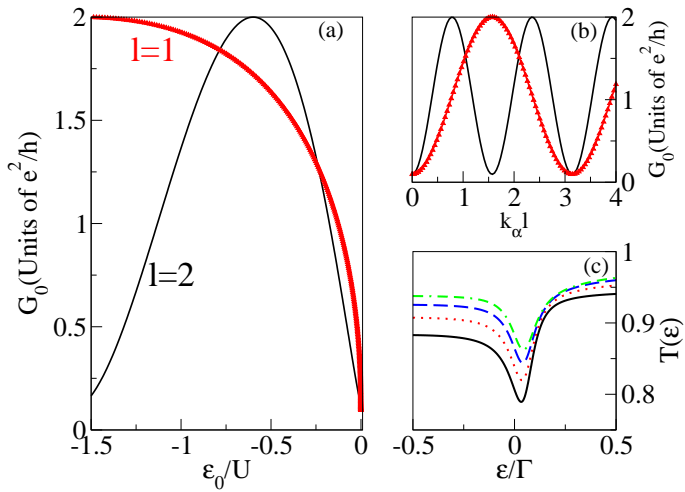}
\caption{(Color online) (a) $\mathcal{G}_{0}$ versus $\epsilon_{0}/U$ in the
  pure Kondo regime for different dot sizes at $k_BT=0$ and $x=0.8$ with
  $E_F=0$. (b) $\mathcal{G}_{0}$ as a function of $\alpha$. (c) $\mathcal{T}(\epsilon)$ for
  different background transmissions (from bottom to top: $x=0.8$, 0.85, 0.9,
  0.95) for $\ell=1$,
$\epsilon_0=-1.5\Gamma$, $U=7\Gamma$ and $k_B
  T=0.05\Gamma$.} \label{fig:3}
\end{figure}

Our results for the Kondo regime in the $k_BT=0$ limit are shown in
Fig.~\ref{fig:3}(a). One might naively think that the ``plateau'' seen in
Fig.~\ref{fig:2} between Coulomb-blockade resonances (which amounts to a
``valley'' in a quantum dot) should decrease due to destructive interference with the
nonresonant path, i.e., 
\begin{equation}
\mathcal{G}_{0}=\frac{2e^2}{h}(1-T_b\cos^2\varphi)\approx 0\,\,\,\,{\rm for}\,\,T_b\approx 1\,,
\end{equation}
as occurs in the AB interferometer.~\cite{bulka01,konig01}
Nevertheless, we see that $\mathcal{G}_{0}$ versus $\epsilon_0/U$ for $\ell=1$
has a strong dependence on the gate voltage. The dependence is stronger for $\ell=2$.
This is due to the fact that $\varphi$ is an implicit function of $\epsilon_0$
through $\alpha$. To confirm this, we plot in Fig.~\ref{fig:3}(b)
$\mathcal{G}_{0}$ as a function of $\alpha$ for values of $\ell=1,2$.
Clearly, $\mathcal{G}_{0}$ in the Kondo plateau is an oscillatory function
of $\alpha$, with the number of oscillations of $\mathcal{G}_{0}$ being
proportional to $\ell$. Finally, in Fig.~\ref{fig:3}(c) we show the total
transmission when $\epsilon_0/\Gamma=-1.5$ for
different background transmissions using the expression of the Rashba dot Green
function given by Eq.~(\ref{GF}) that includes Kondo correlations.
We see that the Kondo effect arises as a Fano-like resonance pinned at
$E_F$ due to its interference with the nonresonant path. As $T_b$ enhances the
Kondo dip is shifted towards higher transmission values. This demonstrates that the $\mathcal{T}$ depends not only on $U$ and $\tilde\Gamma$ but also on
the details of the nonresonant channel via the self-energy.

\section{Conclusions} In short, a quantum wire with a local Rashba field
can sustain a Kondo resonance since a quasibound state emerges from the 
higher subband due to the Rashba interaction.
This state is spin degenerate, it strongly couples with the continuum,
and our results show that considerable
repulsion energy results from charging the dot. 
The conductance resonances in the Coulomb blockade regime
have a Fano form, while in the strong coupling regime we predict an oscillating
$\mathcal{G}_{0}$ as a function of the Rashba strength. We hope that our work
sheds light on the relative influence of spin-orbit and electron-electron
interactions in nanostructures.

\section{Acknowledgments}
We thank M.-S.\ Choi for useful discussions.
This work was supported by the Grant No.\ FIS2005-02796
(MEC) and the Spanish ``Ram\'on y Cajal'' program. 

\end{document}